 \documentclass[letterpaper, 10pt, conference]{ieeeconf}      % Use this line for a4
\IEEEoverridecommandlockouts                              % This command is only
\usepackage{graphicx}
\graphicspath{{../figure/}}
\usepackage{amsmath}
\usepackage{amssymb}
\usepackage{subfig}
\usepackage{mathtools}

\newtheorem{remark}{\textbf{Remark}}

\usepackage{xcolor}
\usepackage{epstopdf}
\usepackage{hyperref}
\usepackage[mathscr]{euscript}

\usepackage{amsfonts, xargs, tensor, units, cite, stmaryrd, mathrsfs, comment}
\usepackage{algorithm}
\usepackage[noend]{algpseudocode}
\usepackage{booktabs}
\usepackage{float}
\newcommand{\HL}[1]{{\color{red}#1}}
\definecolor{orange}{rgb}{1.0, 0.49, 0.0}

\title{A Modular State-Machine Based Event PID Controller}

\author{Sandesh Thapa$^{1}$ and Zhen Qi$^{2}$
\thanks{$^{1}$S. Thapa is with ETAIC Lab, Department of Electrical Engineering,
        University of Texas at Arlington, Arlington, TX, USA,
        {\tt\small sandesh.thapa@uta.edu}}%
\thanks{$^{2}$Z. Qi is a Principal Controls Engineer at Mariana Minerals, TX, USA {\tt\small zhenqi76@gmail.com}}}

\begin{document}
\raggedbottom

\maketitle
\thispagestyle{empty}
\pagestyle{empty}

%%%%%%%%%%%%%%%%%%%%%%%%%%%%%%%%%%%%%%%%%%%%%%%%%%%%%%%%%%%%%%%%%%%%%%%%%%%%%%%%
\begin{abstract}
In this paper, we present a modular proportional-integral-derivative (PID) controller whose computation and mode logic are executed by a higher-level state machine. Inspired by real-time safety-critical applications where computational load, actuator chattering, sensing error and noise are design challenges, the proposed algorithm wraps a standard PID inside a finite state machine with three states (\texttt{pidInit}, \texttt{ErrorOutRange}, \texttt{ErrorInRange}). The goal is to regulate a desired reference within a safe region of operation while reducing actuator chattering and creating a sizable hold band over the range of operation. This design is modular and can be easily integrated into a higher-level state machine with multiple low-level loops and states. The algorithm also has low computational complexity and is suitable for embedded hardware deployment. We demonstrate the effectiveness of the algorithm in simulation on the two test benches of a published event-based PID benchmark \cite{arzen1999event, durand2009further}. The algorithm computes the control updates significantly less often than the time-triggered PID and less often than both dominant event-driven controllers, while keeping a fairly equivalent control performance profile. Furthermore, we also present some experimental results of the designed algorithm for a real-time  flow control.
\end{abstract}

%The algorithm recomputes the control law far less than a time-triggered PID, and less than both published baselines, while keeping a similar regulation profile. We also present experimental results for a control loop running on a real-time system.
%\end{abstract}

%\begin{IEEEkeywords}
%PID control, event-triggered control, state machines, robust control
%\end{IEEEkeywords}

%%%%%%%%%%%%%%%%%%%%%%%%%%%%%%%%%%%%%%%%%%%%%%%%%%%%%%%%%%%%%%%%%%%%%%%%%%%%%%%%
\section{INTRODUCTION}

Proportional-integral-derivative (PID) and its family of controllers are widely used in feedback control, ranging from industrial control to vehicle control and autonomous aerial systems (see e.g., \cite{astrom2006pid, desborough2002increasing, samad2017survey, rajamani2012vehicle, meier2015px4}). Most of these loops are variants of PI or PID. These loops are easy to implement and verify in a low computational cost embedded environment, which is the reason for their wide popularity. These loops often run recursively, and the computations are updated mostly every sample time. While these loops are effective and popular, the continuous sampling-based computation might cause undesired actuator chattering, wear and hysteresis, and decrease the life span of the hardware involved. This can also lead to undesired behavior of the loops.

Is there a fine balance between sampling when needed instead of every sample time? In the early work of {\AA}rz{\'e}n, \cite{arzen1999event} addresses this issue by defining \textit{event-triggered control}. As opposed to sampling at every time instant, this modified version recomputes the control signal only when the error band $e_{lim}$ or maximum sampling time $h_{\max}$ conditions are satisfied. Durand and Marchand \cite{durand2009further} further improve {\AA}rz{\'e}n's algorithm by adding a forgetting factor as a correction term and removing the maximum sampling time $h_{\max}$ condition. These algorithms report a substantial reduction in control law updates. Furthermore, Vasyutynskyy and Kabitzsch \cite{vasyutynskyy2007deadband} analyze deadband sampling in PID control and, in a survey paper \cite{vasyutynskyy2010eventpid}, analyze the family of send-on-delta and deadband formulations. Sanchez et al. \cite{sanchez2011eventpi} compare various event-based sampling strategies on industrial applications. The authors in \cite{beschi2012ssod} define the error as a symmetric staircase and analyze the resulting limit cycles. Similar designs of event-based PID also exist (for e.g., see \cite{vasyutynskyy2007deadband, vasyutynskyy2010eventpid, durand2018quadrotor, durand2015event}).
Our algorithm is closer to the error band holds of \cite{arzen1999event} and \cite{durand2009further}. Thus, in this paper we discuss the comparison with those aforementioned designs.
For bounds on minimum time and Zeno behavior, please see \cite{tabuada2007event, heemels2012eventtriggered, heemels2013petc}. Note that holding the command inside a band is not new, it is a type of deadband PI \cite{vasyutynskyy2007deadband,vasyutynskyy2010eventpid}. %Our contributions are listed below.

In this paper, we design the compute logic that triggers the PID updates as a state machine as opposed to a single equation. Doing so enables integration of higher-level automation and planner state machines into the low-level control seamlessly. One can also design the metric functions to come from higher-level safety or optimization functions. Thus, the design is modular. The PID logic can be in the \textit{ErrorInRange} or \textit{ErrorOutRange} state, and the positions are held during transitions. Error margin and dwell time can be used as design variables to match system requirements.
% We use a constant dwell time $\Delta T$.

The contributions of this paper are as follows. While most event-based PID work is simulation only, we present experimental results of a real-time deployment of an event-based PID controller on embedded hardware. We replace the safety period of \cite{arzen1999event} and the forgetting factor of \cite{durand2009further} with a dwell time $\Delta T$, calibrated from system response, needing no online estimation, and guaranteeing a minimum inter-event time by construction. Our algorithm provides fewer updates than both reference algorithms used for comparison. The update gap grows further under measurement noise. Furthermore, our design is modular and can be easily integrated into higher-level planner or system state machines.
%\HL{We replace the safety period of \cite{arzen1999event} and the forgetting factor of \cite{durand2009further} with a fixed dwell time $\Delta T$. It needs no online estimation, and it guarantees a minimum inter-event time by construction.} 

The rest of this paper is organized as follows. We briefly state the two reference algorithms used as benchmarks in Section~\ref{sec:baselines}. In Section~\ref{sec:design}, we describe our algorithm in detail. We report the simulation study in Section~\ref{sec:results} and the experimental results in Section~\ref{sec:deploy}. Conclusions and future work are discussed in Section~\ref{sec:conc}.
%%%%%%%%%%%%%%%%%%%%%%%%%%%%%%%%%%%%%%%%%%%%%%%%%%%%%%%%%%%%%%%%%%%%%%%%%%%%%%%%

\section{Event-Based PID} \label{sec:baselines}

In this section, we first recall the most commonly used standard and parallel forms of PID controllers. We then briefly provide an overview of two predominant versions of event-based PID controllers. For details, refer to \cite{arzen1999event, durand2009further}. We define the error as $e=r-x$ and let $h_{act}$ denote the time elapsed since the last control update.

\subsection{Standard and Parallel Forms of the PID Controller}

In the \textit{standard form}, the time-based PID controller is defined as:
\begin{equation}
u(t) = K\left( e(t) + \frac{1}{T_i}\int_0^t e(\tau)\,d\tau + T_d\,\frac{d}{dt}e(t) \right)
\label{eq:pid_standard}
\end{equation}

where $T_i$ is the integral time and $T_d$ is the derivative
time. In this form the $K$ gain is applied to
the $I_{out}$ and $D_{out}$ terms. 

In the \textit{parallel form}, the above equation can be rewritten as:

\begin{equation}
u(t) = K_p\,e(t) + K_i\int_0^t e(\tau)\,d\tau + K_d\,\frac{d}{dt}e(t)
\label{eq:pid_parallel}
\end{equation}

where $K_i = K/T_i$ and $K_d = KT_d$. 

\subsection{ {\AA}rz{\'e}n Event-based PID}

The algorithm is briefly stated below for completeness.
\begin{algorithm}[H]
\caption{{\AA}rz{\'e}n's event-based PI \cite{arzen1999event}}
\label{alg:arzen}
\begin{algorithmic}[1]
\State $e \gets r-x$; \quad $h_{act} \gets h_{act} + d\tau$
\If{$|e-e_{\mathrm{old}}| > e_{lim}$ \textbf{ or } $h_{act} \ge h_{\max}$}
  \State $u \gets K e + u_I$
  \State $u_I \gets u_I + \tfrac{K}{T_i}\,h_{act}\,e$
  \State $e_{\mathrm{old}} \gets e$; \quad $h_{act} \gets 0$
\EndIf
\end{algorithmic}
\end{algorithm}

\subsection{The Algorithm of Durand and Marchand}
Durand and Marchand \cite{durand2009further} improve {\AA}rz{\'e}n's algorithm by adding a forgetting factor as a correction term and removing the maximum sampling time $h_{\max}$ condition from the first condition and making it nested. The update law becomes: 
\begin{equation}
u_I \leftarrow u_I + \tfrac{K}{T_i}\,h_{act}\,e
\label{eq:accum}
\end{equation}
Out of the four algorithms presented by Durand and Marchand \cite{durand2009further}, it is reported that the hybrid algorithm performs the best. We refer the reader to \cite{durand2009further} for a detailed overview.

\begin{algorithm}[H]
\caption{Hybrid algorithm of Durand and Marchand \cite{durand2009further}}
\label{alg:dm}
\begin{algorithmic}[1]
\State $e \gets r-x$; \quad $h_{act} \gets h_{act} + d\tau$
\If{$|e-e_{\mathrm{old}}| > e_{lim}$} %\Comment{no $h_{\max}$ term}
  \If{$h_{act} \ge h_{\max}$}
    \State $h_{act}^{\,i} \gets h_{act}\exp(d\tau-h_{act})$ % \Comment{forgetting}
    \State $he \gets (h_{act}^{\,i}-d\tau)\,e_{lim} + d\tau\,e$ % \Comment{bound}
  \Else
    \State $he \gets h_{act}\,e$
  \EndIf
  \State $u_I \gets u_I + \tfrac{K}{T_i}\,he$; \quad $u \gets K e + u_I$
  \State $e_{\mathrm{old}} \gets e$; \quad $h_{act} \gets 0$
\EndIf
\end{algorithmic}
\end{algorithm}

%%%%%%%%%%%%%%%%%%%%%%%%%%%%%%%%%%%%%%%%%%%%%%%%%%%%%%%%%%%%%%%%%%%%%%%%%%%%%%%%
\section{Control Design} \label{sec:design}

In this section, we state our proposed algorithm in detail.

\subsection{Normalized Error} \label{sec:normerr}

We define the normalized error as:
\begin{equation}
e_n(t_k) :=\frac{r(t_k)-x(t_k)}{r(t_k)},
\qquad e_n^{per}(t_k)=100\,e_n(t_k),
\label{eq:err}
\end{equation}
and the \textit{ErrorInRange} band is defined as: 
\begin{equation}
e^{\mathrm{inRange}}(t_k)=\bigl[\,|e_n^{per}(t_k)|\le(e_n^{per})_{des}\,\bigr].
\label{eq:inrange}
\end{equation}

Note that the switching variable in our design is expressed as a percentage. Normalizing the error band with respect to the setpoint decouples the band from the absolute scale of the setpoint. This is particularly useful where the setpoint and operating regime have a wide envelope. We understand that this design makes the gains setpoint dependent. However, it also makes the gains and margin intuitive for tuning. It also matches the way sensor and instrument tolerances are specified. Further, learning-based tools will be explored to address this issue in future research. The controller is defined on an operating envelope $r(t_k)\in[r_{\min},r_{\max}]$ with $r_{\min}>0$. For the case where $r_k \to 0$, the denominator is clamped at $r_{\epsilon}>0$ to prevent a singularity. We set $r_{\epsilon}=10^{-4}$ in both simulation and experimental sections.

\subsection{System Level Design}

\begin{figure}[H]
  \centering
  \includegraphics[width=\columnwidth]{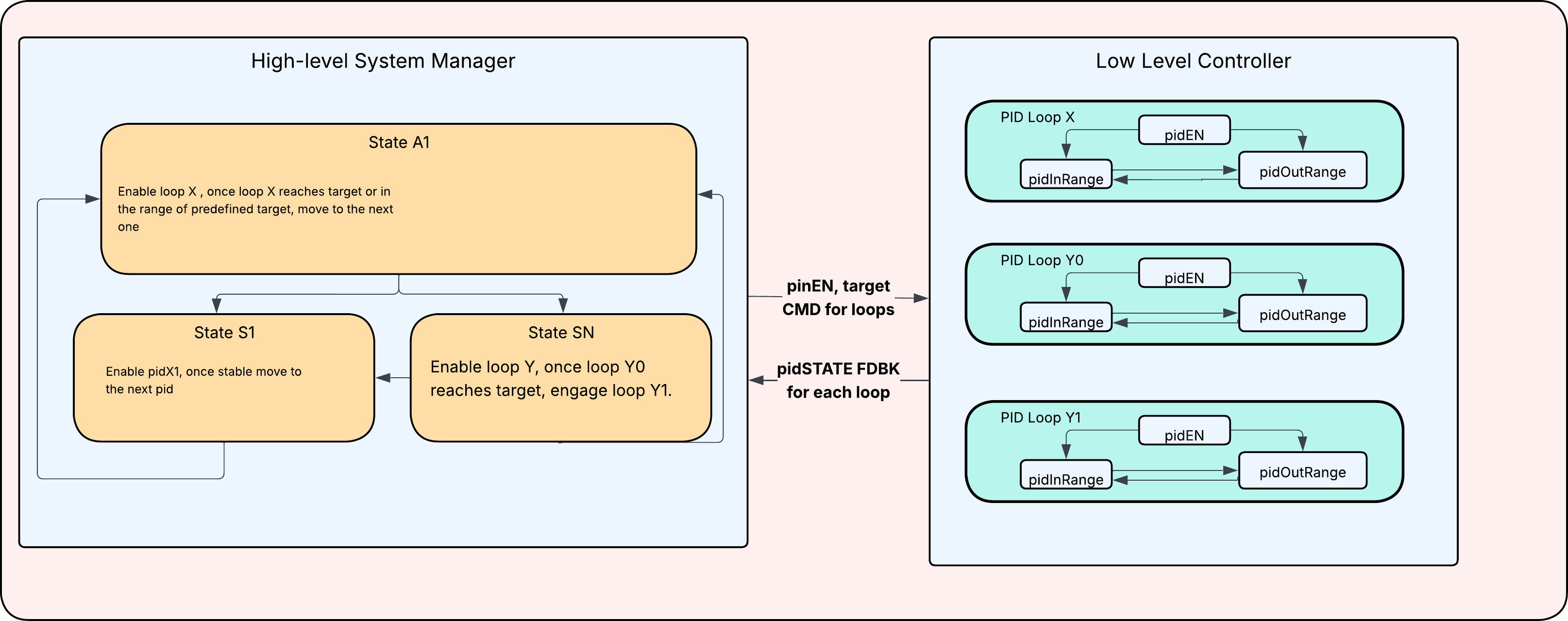}
  \caption{An outline of how modular PID can be integrated into higher-level automation tasks.}
  \label{fig:sysdesign}
\end{figure}

The integration of the modular design with higher-level autonomy is shown in Fig.~\ref{fig:sysdesign}.

\subsection{State Machine Based PID}

\begin{figure}[H]
  \centering
  \includegraphics[width=\columnwidth]{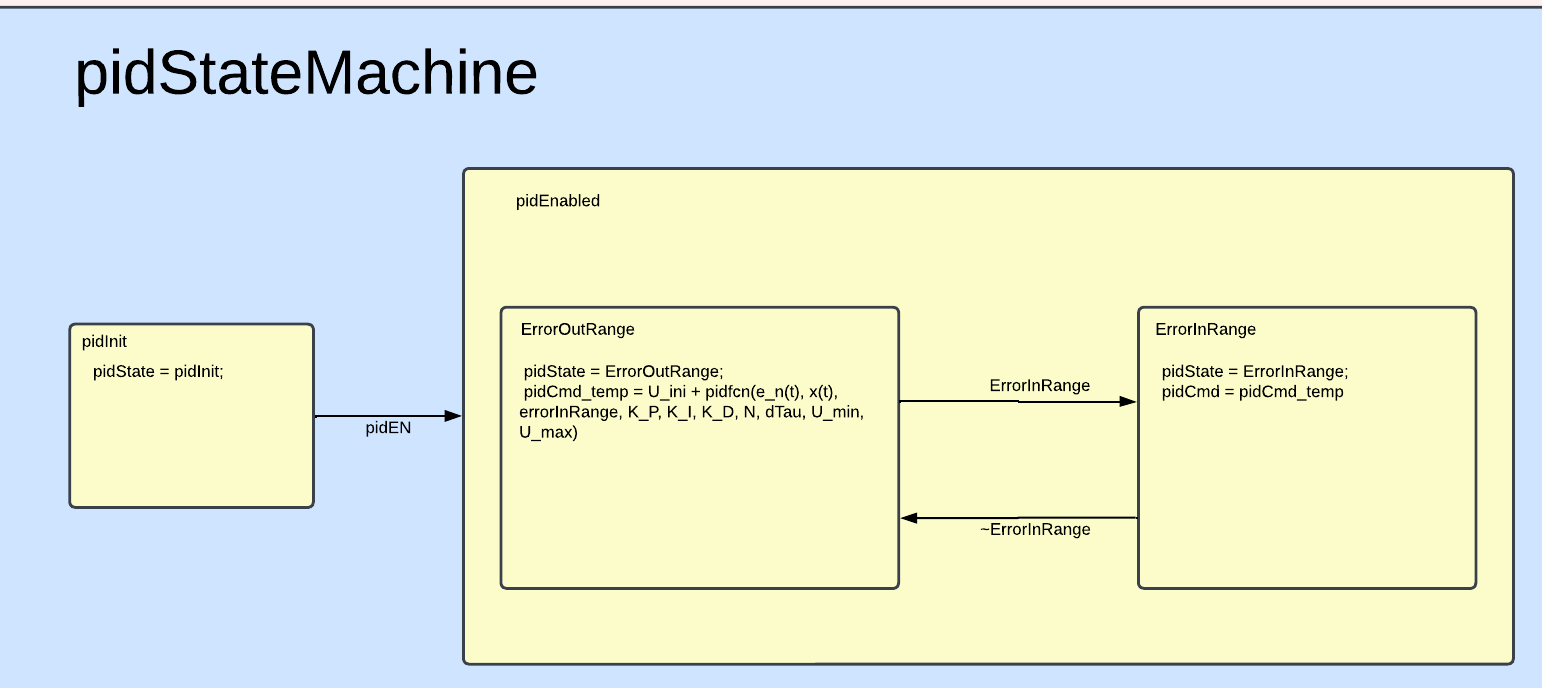}
  \caption{Detailed view of PID Control with its sub-states.}
  \label{fig:chartzoom}
\end{figure}

Fig.~\ref{fig:chartzoom} shows the internal \texttt{pidStateMachine}, with \texttt{pidInit} and the \texttt{ErrorOutRange}/\texttt{ErrorInRange} sub-states of \texttt{pidEnabled}.

The proposed state-machine PI algorithm is stated next.

\begin{algorithm}[H]
\caption{Proposed state-machine PI}
\label{alg:chart}
\begin{algorithmic}[1]
%\Require $(e_n^{per})_{des}$, $\Delta T$, $K_P$, $K_I$, $U_{ini}$
\State $u_I(0) \gets 0$; \quad $u_{mid}(0) \gets 0$
%\For{every base sample $k$}
  \State $e_n(k) \gets (r(k) - x(k))/r(k)$
  \If{$|e_n^{per}(k)| > (e_n^{per})_{des}$}
      \State \textit{pidState} $\gets$ \textit{ErrorOutRange}
      \State $u_{mid}(k) \gets K_P e_n(k) + u_I(k-1)$
      \State $u_I(k) \gets u_I(k-1) + K_I e_n(k)$
      \State wait ( $\Delta T$,  \textit{sec})
  \Else
      \State \textit{pidState} $\gets$ \textit{ErrorInRange}
      \State $u_I(k) \gets u_I(k-1)$
      \State $u_{mid}(k) \gets u_{mid}(k-1)$
  \EndIf
  \State $u_{unsat}(k) \gets U_{ini} + u_{mid}(k)$
  \State $u_{sat}(k) \gets \mathrm{sat}(u_{unsat}(k), U_{min}, U_{max})$
%\EndFor
\end{algorithmic}
\end{algorithm}

\begin{remark}[PID vs. PI, and Zeno-freeness]
Throughout this paper, $K_D=0$. Our controller runs as PI\HL{,} which is consistent with \cite{arzen1999event,durand2009further}. The derivative term can be introduced if needed. $\Delta T > 0$ fixes the minimum dwell by construction. Thus, our design prevents Zeno behavior with no safety period or forgetting factor needed.
\end{remark}

\section{Simulation Results} \label{sec:results}
We present four different simulation cases to validate our algorithms. 
First, we validate our design with two established benchmarks of {\AA}rz{\'e}n \cite{arzen1999event} and Durand and Marchand \cite{durand2009further} (Algorithms \ref{alg:arzen} and \ref{alg:dm}). We also provide simulation results for a measurement noise sweep and time-varying reference tracking for all four algorithms, including the standard time-based PID, along with our design (Algorithm \ref{alg:chart}). The results are reported in Table \ref{tab:bench}.

We adapted the plant model and gains used in \cite{durand2009further}. The model is given by $H(s)=1/(1+s)$ with a zero-order hold, simulated for $20$~s \cite{durand2009further}, where the gains used are $K=1.83$, $T_i=0.457$, $h_{nom}=0.05$~s, $h_{\max}=0.5$~s, and
$e_{lim}=0.01$. Since the state-machine PI is developed in parallel form, $K_I=(K/T_i)\Delta T=0.2002$, and we set $\Delta T=h_{nom}$.

\begin{remark}[Steady-state accuracy and limit-cycle risk]
Steady-state accuracy converges to the calibrated band ($0.5\%$ in simulation, $5.0\%$ in the deployment). This is a known tradeoff of such deadband controllers \cite{vasyutynskyy2007deadband,vasyutynskyy2010eventpid}. Such deadband with integral action can also lead to limit cycles on certain systems \cite{choudhury2005stiction,tao1996adaptive}.
For the experiments reported here, no such behavior was observed. We did not run a dedicated limit-cycle test. We plan to investigate that further in future studies. 
\end{remark}

\subsection{Setpoint-Dependent Gain Scaling} \label{sec:scaling}
In order to be consistent with the simulation benchmark, we now briefly state the gain scaling used in the simulation sections:

%Because the chart acts on the normalized error $e_n^{per}=(r-x)/r$ of Section~\ref{sec:normerr}, a fixed gain pair $(K,K_I)$ gives an effective gain on the raw error of $K/r$, which falls as the setpoint moves above the point the gains were tuned at. $K_P$ and $K_I$ are scaled with the setpoint so that the effective gain, and therefore the comparison against the reference algorithms, stays consistent across Bench~1, Bench~2, and every other case:
\begin{equation}
K_P(new) = K_P\!\left(\frac{r}{r_{ref}}\right), \qquad
K_I(new) = K_I\!\left(\frac{r}{r_{ref}}\right),
\label{eq:kp}
\end{equation}

\subsection{Case I: Setpoint Steps (Bench 1)}
We reproduced the simulation case of \cite{durand2009further} and
\cite{arzen1999event}. The reference is defined as:

\begin{equation}
r(t) =
\begin{cases}
0, & 0 \le t < 1 \\
1, & 1 \le t < 10 \\
2, & t \ge 10
\end{cases}
\label{eq:testcase1}
\end{equation}

For the state-machine PI, we set $(e_n^{per})_{des}=0.5\%$.

\begin{table}[H]
\caption{Control updates out of $400$ nominal samples, bench 1 / bench 2.}
\label{tab:bench}
\centering
\begin{tabular}{@{}lcc@{}}
\toprule
Controller & Control Update & Control Update \% \\
\midrule
Time-triggered PI                                    & $400/400$ & $100.0\%/100.0\%$ \\
{\AA}rz{\'e}n, with $h_{\max}$ \cite{arzen1999event} & $101/82$  & $25.3\%/20.5\%$  \\
Hybrid \cite{durand2009further}                      & $67/43$   & $16.8\%/10.8\%$   \\
State-Machine PI                                     & $\mathbf{55/33}$ & $\mathbf{13.8\%/8.3\%}$ \\
\bottomrule
\end{tabular}
\end{table}

\begin{figure}[H]
\centering
\includegraphics[width=0.85\columnwidth]{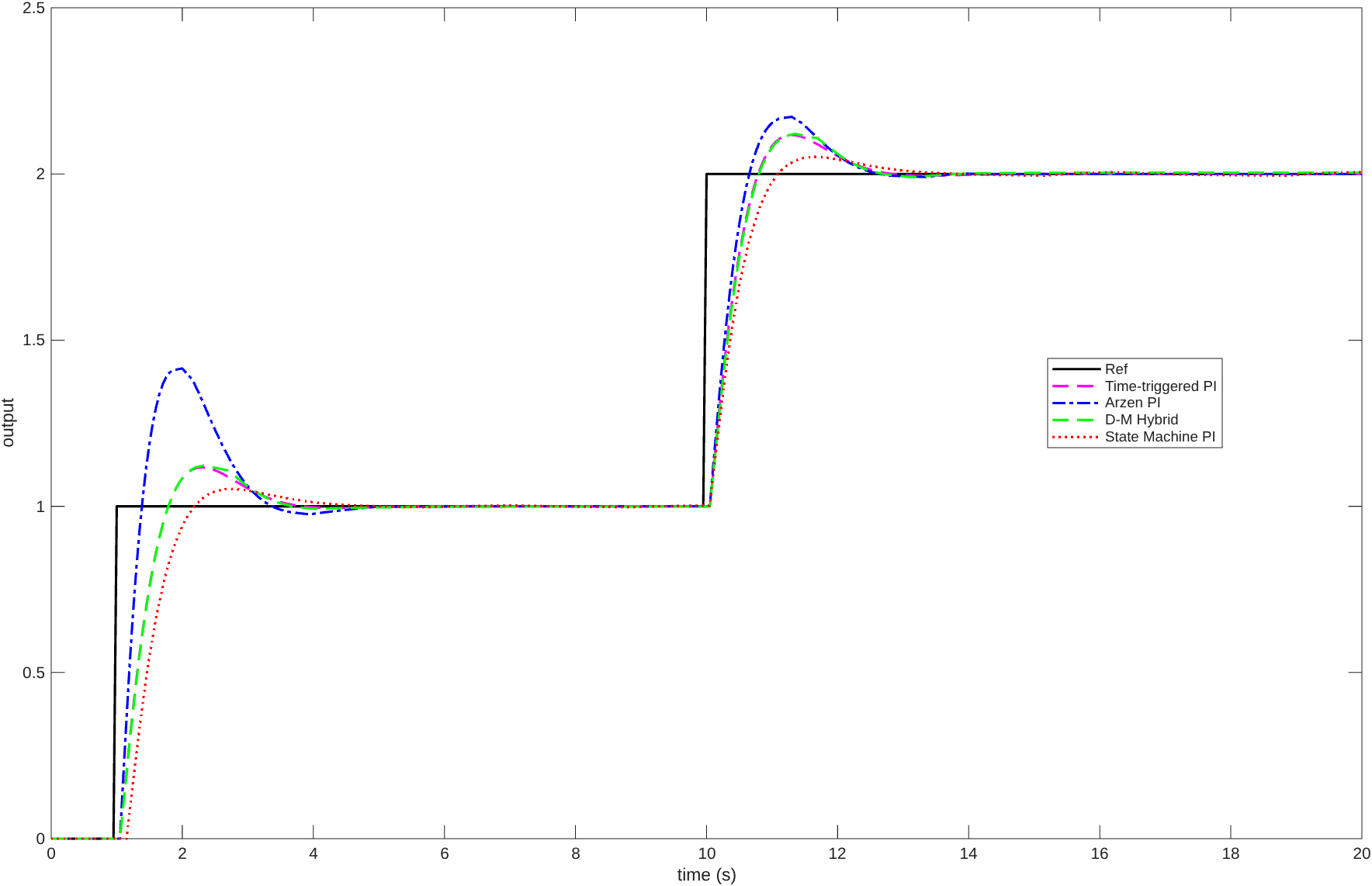}
\caption{ Response of each controller with setpoints for Case I. Our proposed algorithm tracks with $\HL{\sim}5.0\%$ overshoot and a $\HL{\sim}0.70$~s rise time as shown above. Detailed comparison is reported in (Table~\ref{tab:bench})}
\label{fig:bench1}
\end{figure}

As shown in Fig.~\ref{fig:bench1}, the proposed controller tracks the step command with $\sim5.0\%$ overshoot and a $\sim0.70$~s rise time. The hybrid algorithm of Durand and Marchand \cite{durand2009further} has $\sim12\%$ overshoot and {\AA}rz{\'e}n's has $\sim41\%$. The rise time of $\sim0.70$~s is much slower than the $\sim0.50$~s of {\AA}rz{\'e}n's design, which is an expected result because of our hold band.

\subsection{Case II: Setpoint Step with Load Disturbance (Bench 2)}
In this simulation, we resimulated the load-disturbance case of \cite{durand2009further}, where the reference is defined as:
\begin{equation}
r(t) =
\begin{cases}
0, & 0 \le t < 2 \\
1, & t \ge 2
\end{cases}
\qquad
d(t) =
\begin{cases}
0, & t < 12 \\
0.1, & t \ge 12
\end{cases}
\label{eq:testcase2}
\end{equation}

The disturbance is added to the input load at $12$~s.

\begin{figure}[H]
\centering
\includegraphics[width=0.85\columnwidth]{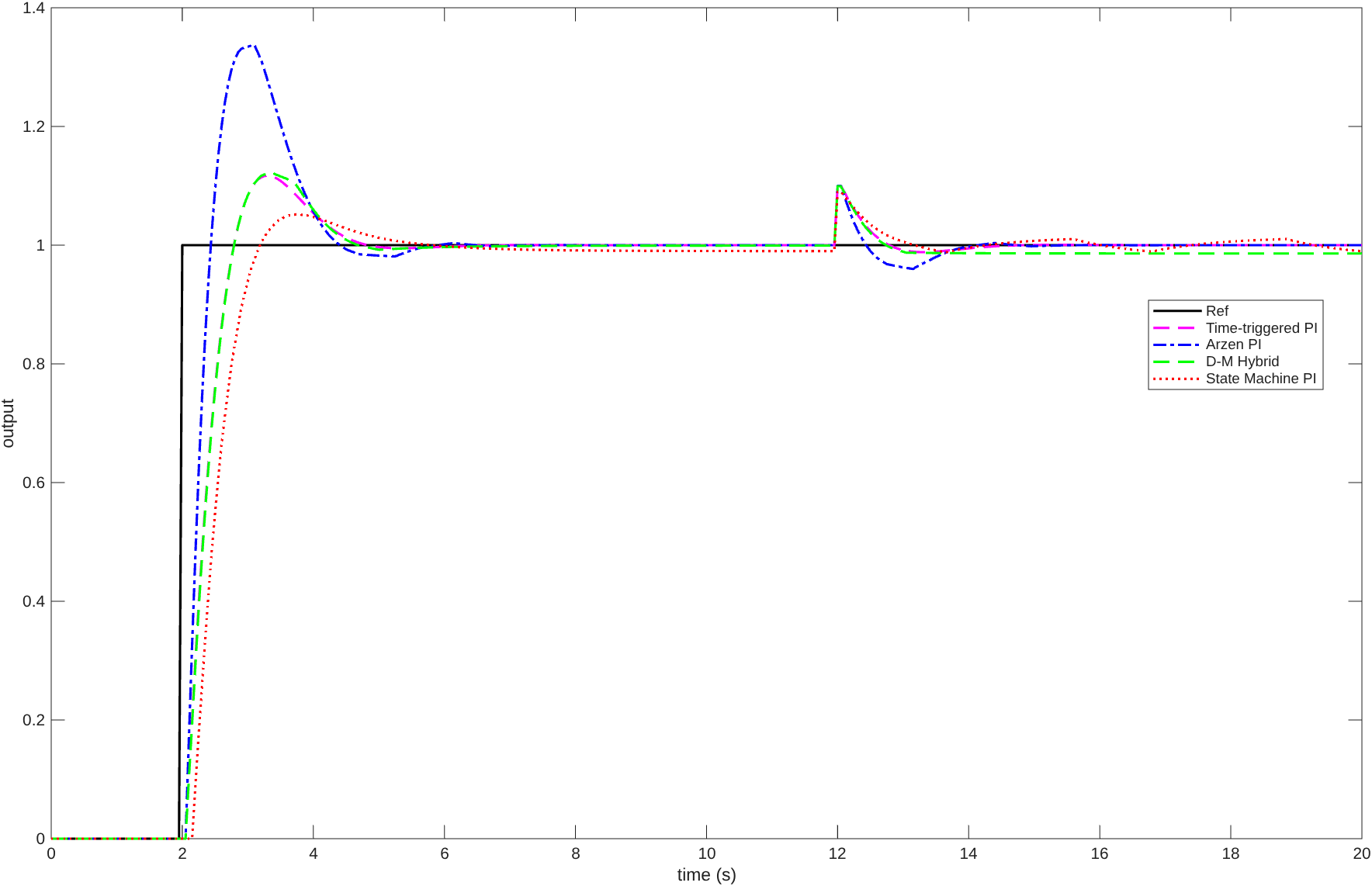}
\caption{
 Response of each controller with setpoints and load disturbance for Case II. Our proposed algorithm has similar performance to Case I and recovers with fewer updates. Detailed comparison is reported in (Table~\ref{tab:bench})}
\label{fig:bench2}
\end{figure}

Fig.~\ref{fig:bench2} shows that the proposed controller recovers from a disturbance with residual error lower than the hybrid algorithm. It also reports a lower update count, as shown in Table~\ref{tab:bench}.

\subsection{Case III: Measurement Noise Sweep}

We further compare the four designs with a noise test. We used band-limited white noise as an additive disturbance to the measurement to simulate more realistic scenarios.
\begin{equation}
y_{meas}(t) = y(t) + n(t), \qquad n(t) \sim \mathcal{N}(0,\sigma^2)
\label{eq:noise_meas}
\end{equation}
\begin{equation}
\sigma \in \{\sigma_1, \sigma_2, \ldots, \sigma_M\}
\label{eq:noise_levels}
\end{equation}

We fix $(e_n^{per})_{des}=1\%$
and vary $\sigma$ over $\{0,\,0.002,\,0.005,\,0.01,\,0.02\}$.

The reference is defined as: 
\begin{equation}
r(t) =
\begin{cases}
0, & 0 \le t < 2 \\
1, & t \ge 2
\end{cases}
\label{eq:ref_case3}
\end{equation}

\begin{figure}[H]
\centering
\includegraphics[width=0.85\columnwidth]{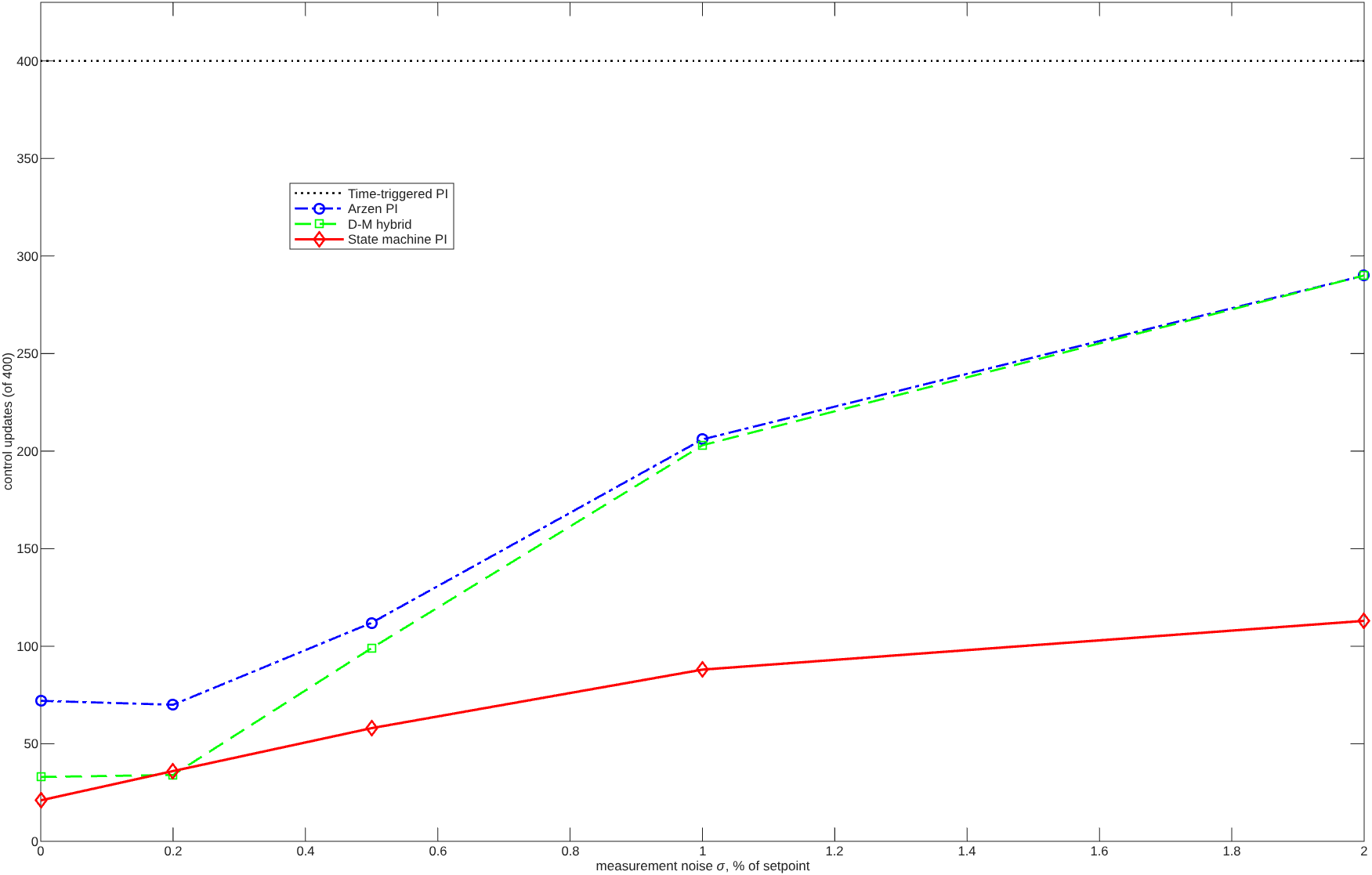}
\caption{ Control updates against a measurement noise sweep while all controllers keep $(e_n^{per})_{des}=1\%$. Our design's update count grows slower than the other compared algorithms.}
\label{fig:sigma}
\end{figure}

Fig.~\ref{fig:sigma} shows that the proposed controller's update count grows much more slowly as band-limited white noise is introduced. We kept the rest of the parameters constant throughout this case study. At the highest noise level simulated, we issued roughly $113$ updates as compared to roughly $290$ updates for {\AA}rz{\'e}n and the hybrid algorithm of Durand and Marchand.

\begin{figure}[H]
\centering
\includegraphics[width=0.85\columnwidth]{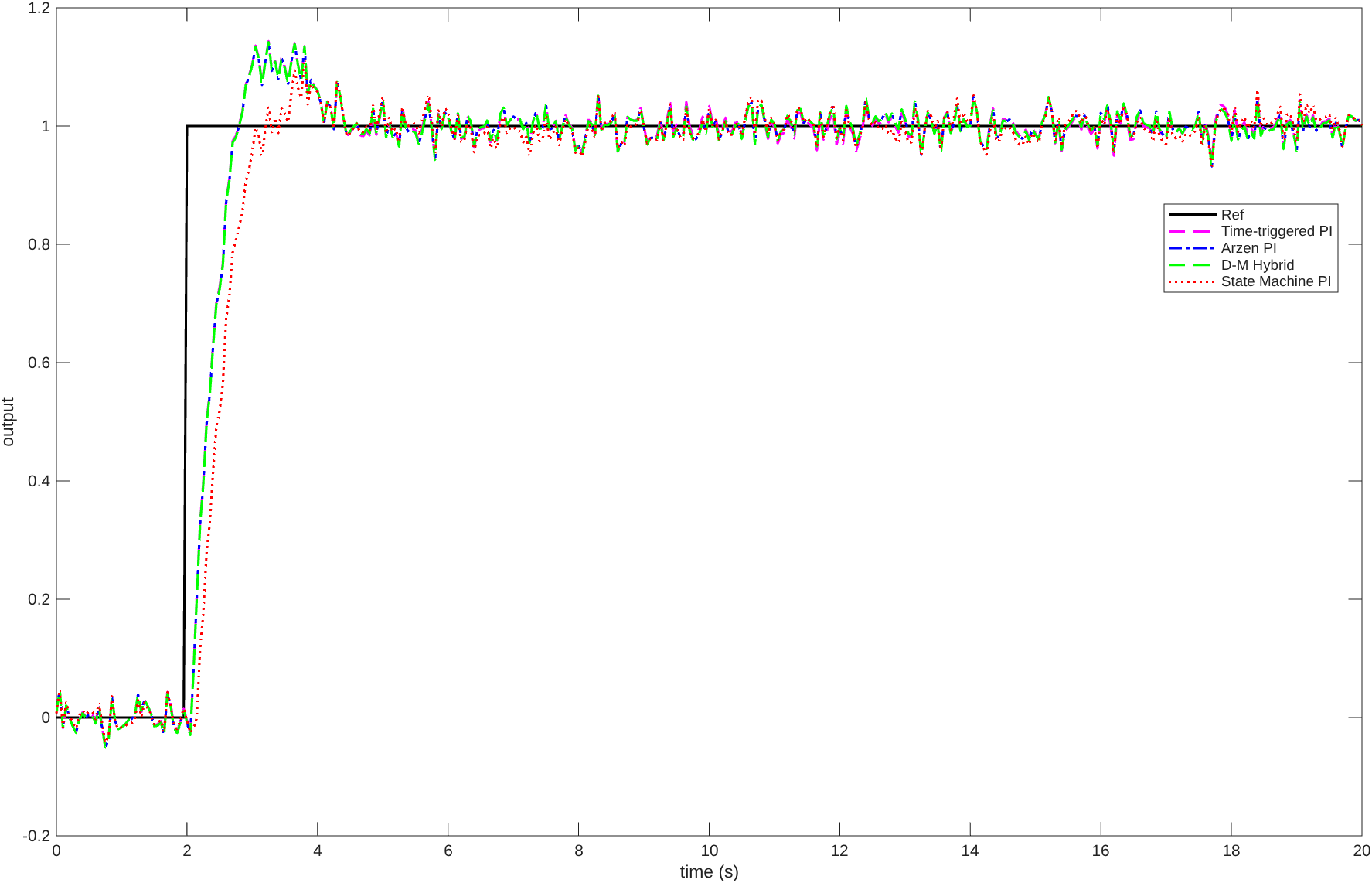}
\caption{ Closed-loop response of all four algorithms under measurement noise for $\sigma=0.005$. All the controllers have similar tracking behaviors, while updates on our design remain relatively low.}
\label{fig:sigma005}
\end{figure}

We also present the tracking profile for the noise sweep for the case when $\sigma=0.005$, as depicted in Fig.~\ref{fig:sigma005}. As stated, the tracking profile remains comparable across the sweep, and the proposed controller updates less. It can be noted that this design parameter is a tradeoff: it can be used to obtain comparable tracking behavior with lower updates for the control law, which helps in extending the life of the actuator and preventing chattering.

\subsection{Case IV: Time-Varying Setpoint}

We further evaluate the performance of our algorithms with a time-varying setpoint. The other parameters are kept
unchanged from previous simulations.
We define the reference as:

\begin{equation}
r(t) = \begin{cases}
0.5 + 0.625\,t, & t \le 4~\text{s},\\
3, & 4 < t \le 16~\text{s},\\
3 - 0.625\,(t-16), & 16 < t \le 20~\text{s}.
\end{cases}
\end{equation}

\begin{figure}[H]
\centering
\includegraphics[width=0.75\columnwidth]{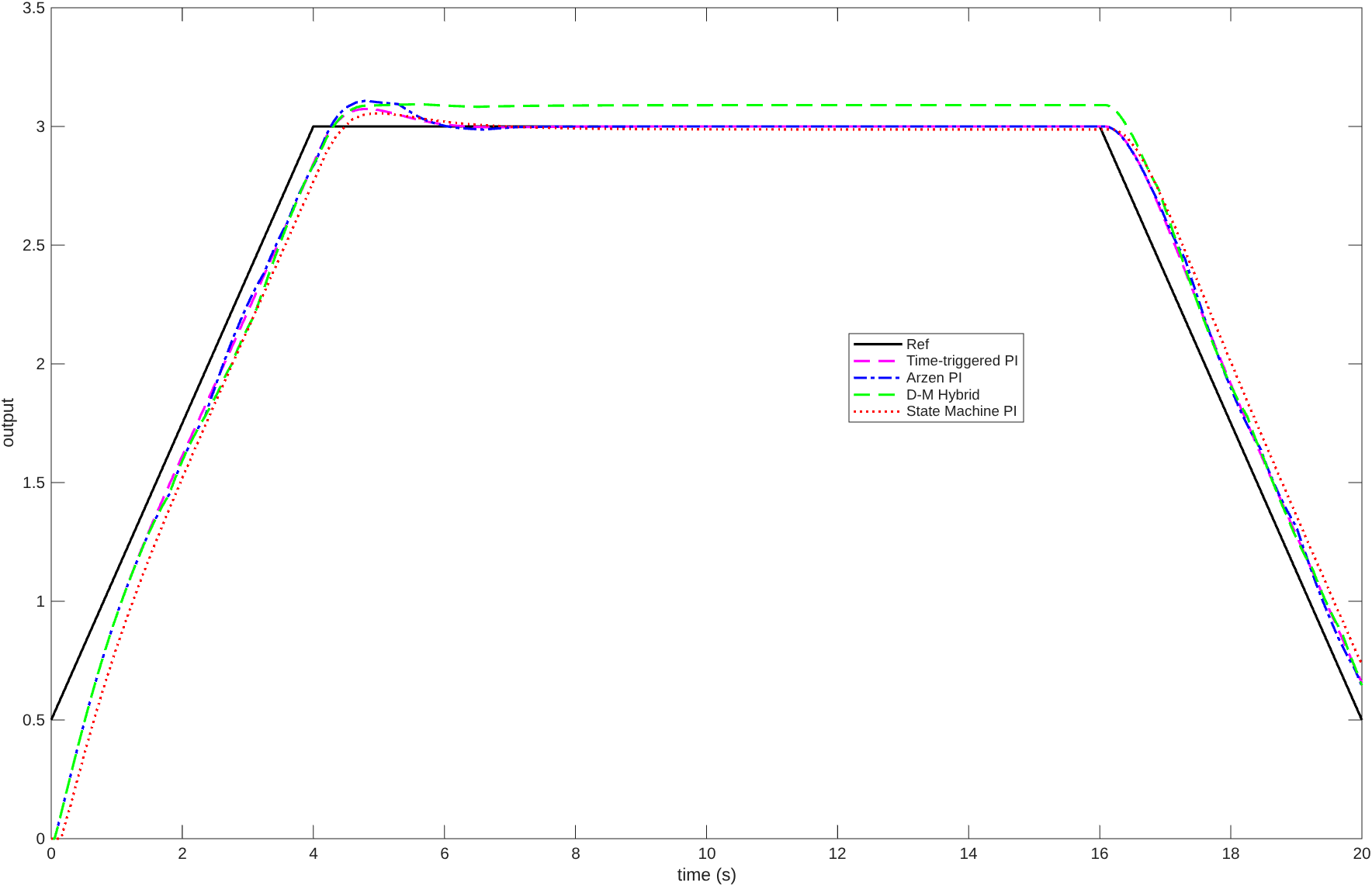}
\caption{Controller performance against a time-varying setpoint. The tracking profile remains similar to the other event-based controllers, and it issues fewer updates.}
\label{fig:trapsched}
\end{figure}

Fig.~\ref{fig:trapsched} shows that the proposed controller tracks a time-varying setpoint. The overshoot remains negligible, and the controller is triggered less than the others. The update counts are $68$ for our algorithm versus {\AA}rz{\'e}n's $96$ and the hybrid algorithm's $74$. This case study further shows that, without compromising tracking quality, we are able to reduce the controller update counts.

\section{Experimental Results} \label{sec:deploy}
We validate the performance of the proposed state-machine PI controller on an experimental platform. We present a representative case study from our experiments. Our system is a flow control loop with volumetric flow regulated through a control valve. 
% \HL{Our test system is an airflow control loop, with volumetric flow regulated through a control valve.} 
 The experimental platform consists of a 32-bit Ecotron EV22297A03 VCU and a Vector CANape HMI environment for gain tuning and calibration. We generated the state machine chart to C. The controller occupies less than $100$ bytes. For our experimental platform, we set $K_P=40$, $K_I=0.05$, $U_{ini}=15.5$, $[U_{min},U_{max}]=[0,100]$, $(e_n^{per})_{des}=5\%$ and $\Delta T=0.3$~s. Note that the set of gains in simulation is different from the experimental platform. The gains provided in \cite{durand2009further}, $K=1.83$, $T_i=0.457$, come from a reduced-order model with a plant of $H(s)=1/(1+s)$. Thus, these gains are only valid for that specific simulated plant. Our experimental gains, $K_P=40$, $K_I=0.05$, were tuned during on-site calibration and testing with a step response. We also did not have access to a quantitative model, so we relied on field tuning of these gains.

\begin{figure}[H]
\centering
\includegraphics[width=0.9\columnwidth]{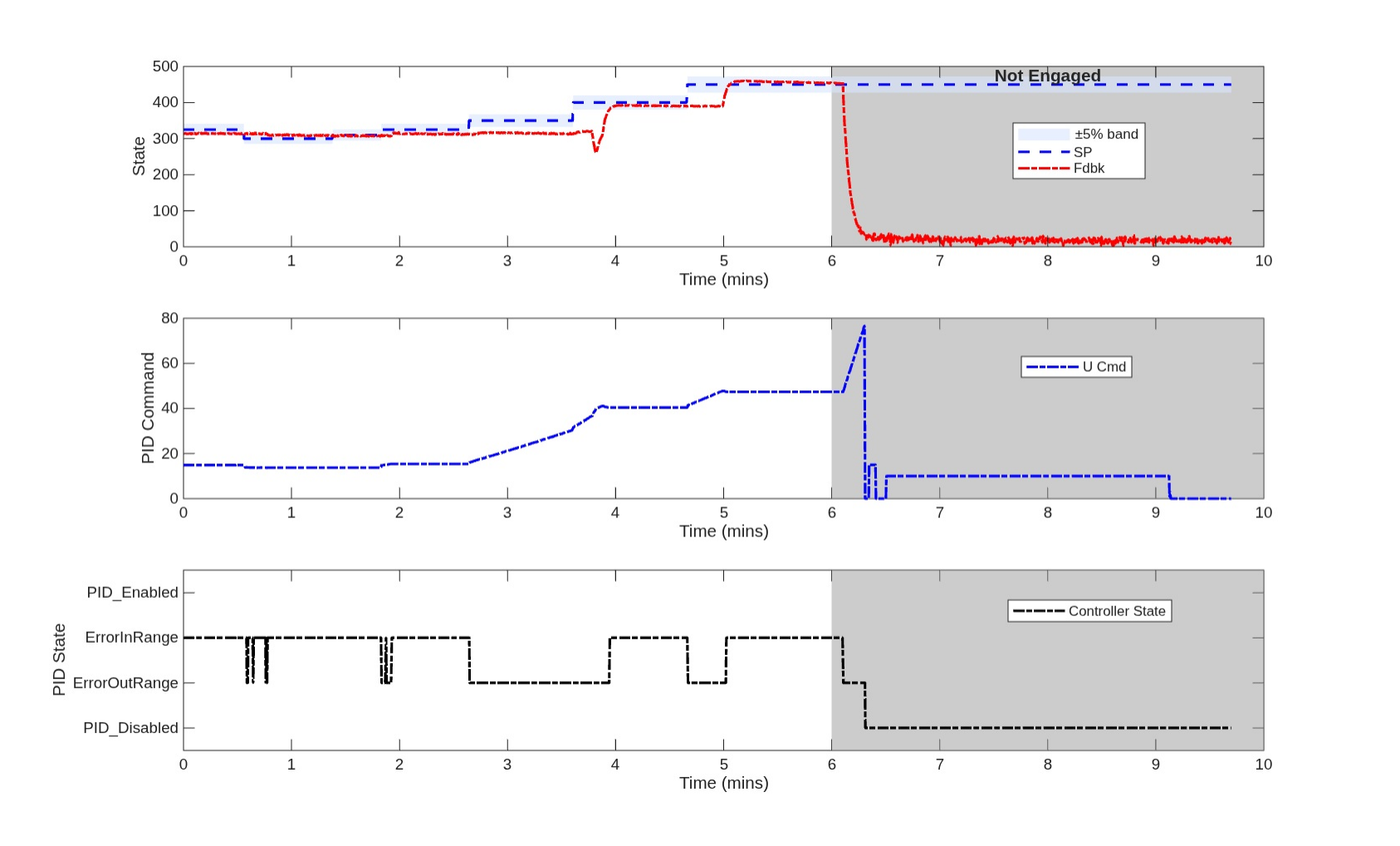}\caption{Setpoint, measured state, hold band, PID command, and PID state trajectory for an experimental case study.}
\label{fig:fullrun}
\end{figure}

The control element is a control valve which are known to be prone to deadband problems \cite{coughran1998deadband}. We left the valve model, flow range and duct sizing out for proprietary reasons. We put the controller in automatic mode from $t=0.00$ to $t=6.31$~min. After that, we manually control the system and turn off the system. We limit the discussion for the rest of the experimental section and report the tracking performance and update counts for $t\in[0,6.31]$~min. The sample time is $0.1$~s.
We further conducted the sensor noise analysis and report the following findings:

\begin{itemize}
\item We estimate the noise from six windows where both the setpoint and the command were held constant for at least $30$~s. These include setpoints of $300$, $310$, $325$, $400$, and $450$.
\item Across these windows, the average standard deviation was measured to be  ~$0.4\%$ from the setpoint.
\item The noisiest window had an average standard deviation of ~$0.7\%$ of the setpoint for $450$ of flow setpoint.
\item We found the largest deviation to be around $4.62\%$ for the above reported windows.
\item As a safe range, we keep our hold band to be $5\%$ to account for worse case scenarios. This is done for safety for higher operating regimes. 
\end{itemize}
Fig.~\ref{fig:fullrun} shows one full experimental run. We report the following findings computed from the experimental log over $t\in[0,6.31]$~min:
\begin{itemize}
\item The controller recomputes $389$ times out of $3{,}785$ engaged samples. This is about $10\%$ of what a time-triggered control would issue updates. 
\item Actuator command variation is about $1.0\%$/min during holding. And about $20.0\%$/min while it is controlling. This implies roughly a $20\times$ reduction which helps in reducing actuator chattering. For a system running hundreds of hours, this is a significant reduction in action. Hence contributing to higher life spans of such actuators.
\item Throughout the experiments, we show that tracking is also preserved as shown by the steady state error converging inside the band defined. 
%About $70\%$ of \texttt{ErrorInRange} samples are inside the $\pm5\%$ band\HL{ [Flag: reporting accuracy only within \texttt{ErrorInRange} is near-circular, since that state is defined by this same band -- consider reporting \% of all engaged samples instead].}
\end{itemize}

\begin{remark}
We did not run a dedicated benchmark test for the experiments with another event-based PI due to lack of further testing facility. We plan to add that to future work once we have access to experimental platforms again.
\end{remark}

\begin{figure}[H]
\centering
\includegraphics[width=0.9\columnwidth]{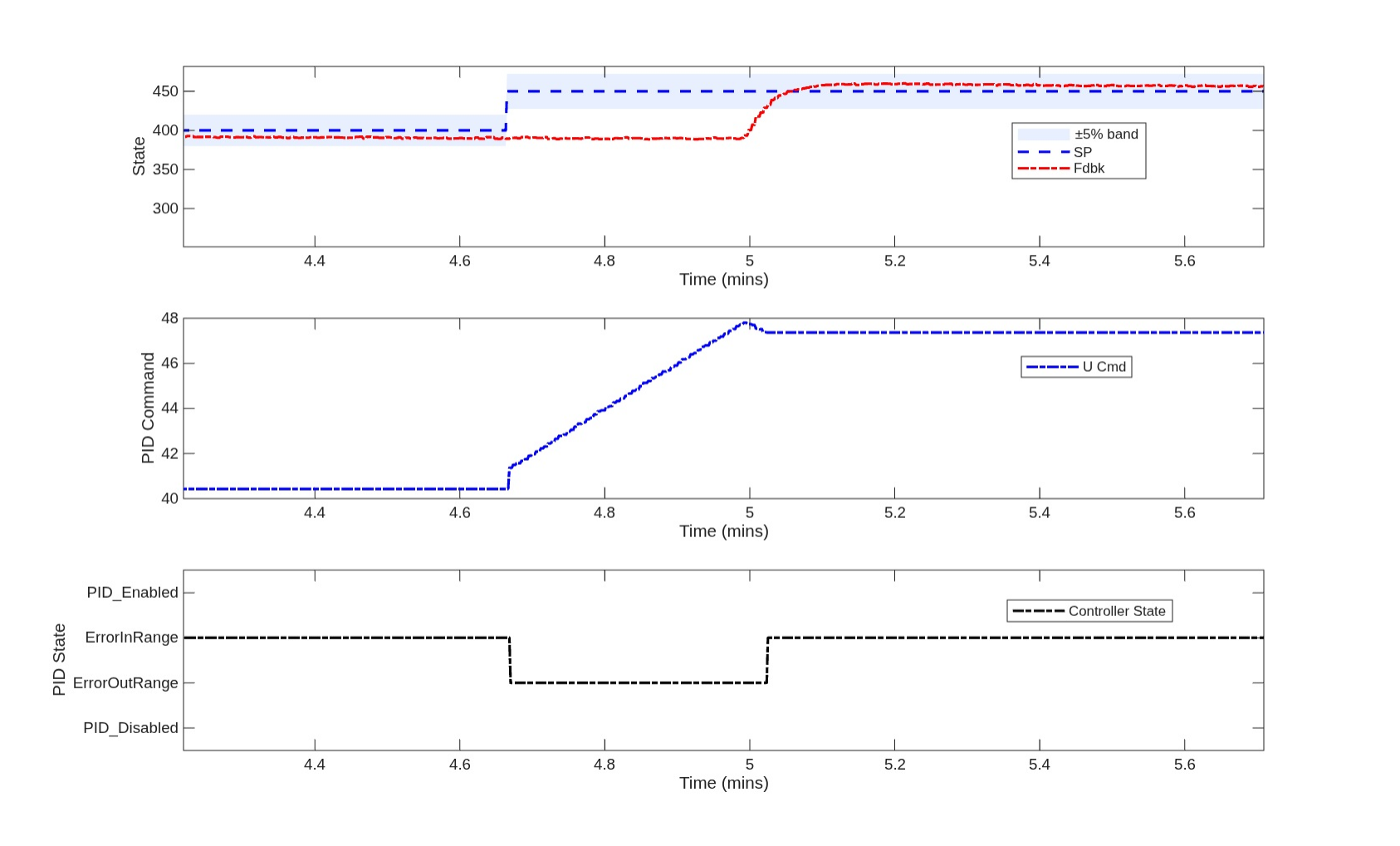}\caption{A zoomed-in snapshot of setpoint, measured state, hold band, PID command, and PID state.}
\label{fig:step}
\end{figure}

In Fig.~\ref{fig:step}, we show how the PID command is held while the PID state is in range. Once the hold ends, the command is updated from the previous step. We also discovered that our system has a fair amount of lag because of the placement of sensors. The valves used also had around $1.0\%$ of deadband \cite{choudhury2005stiction, coughran1998deadband}. Thus, we tuned our system for a smooth transient as opposed to the fastest recovery. Given the proprietary constraint, we keep the literature range of deadband to be less than $1.0\%$. This number can be easily discovered once you have the data sheets and calibrations from specific vendors.


\subsection{Lessons Learned}

\paragraph{Gain Tuning and Calibration}
Some practical approaches used to find a good hold band $(e_n^{per})_{des}$
and update interval $\Delta T$ are briefly stated next. One standard procedure used in testing is to take the
minimum system error reported by the sensor as the hold band. One can also apply
multiple constant commands and check the deviation in the measurements to find the
minimum error resolution in the loop. The starting error percent can also be taken
from the sensor data sheet and further calibrated from field tests. The error range
is only changed if the operating dynamics change and tighter tracking is required.
Similarly, $\Delta T$ is kept constant most of the time, but can be changed if the
operating regime changes. Since the system lacks a model, $\Delta T$ is used as a
tuning factor when the system has a large lag between the sensor and the command
output. Note also that the error is normalized so that the PI gains remain
realistic during tuning for a real-time system with a large operating range.

\paragraph{Gain Transfer from Simulation}

Since our current system lacked a deterministic model, we could not transfer the gains from the simulation benchmarks to the experiments. The simulation benchmarks are shown here to compare our approach against the established benchmarks. One can also tune gains in a high-fidelity simulation model, such as Simscape \cite{mathworks2024simscape} or Modelon \cite{modelon2024impact}.

\paragraph{Deadband Tuning}

For actuators, one can always tune the system to have a minimum deadband, but note that because of the non-linear nature of DC and stepper motors, this non-linearity is part of the design. One should avoid those regimes where the behaviors of actuators are not well-defined. This deadband causes hysteresis and thus contributes to sensor error, which impacts our hold band \cite{tao1996adaptive}.

\section{Conclusion and Future Work} \label{sec:conc}

In this paper, we present a modular state-machine based PID controller. Our trigger condition is a state defined by an error band rather than a single compensation equation. We provided simulation results with three other benchmarks and provided a comparison of tracking performance. Our controller kept a similar tracking profile while reducing updates. We also presented a case of a noise band sweep and a time-varying reference trajectory and analyzed the performance. Our controller matches the benchmarks and reduces the update count. Furthermore, we presented experimental results of a real-time control loop. Future work will focus on adding learning-based algorithms \cite{baumann2018deep, boffi2021learning}
for learning the error band and dwell time. We also plan to investigate data-driven safety certificates \cite{ames2019cbf,
breeden2022sampled}. We further plan to extend this algorithm to UAV trajectory tracking and vehicle speed control problems.

\bibliographystyle{IEEEtran}
\bibliography{IEEEabrv,mybib}

\end{document}